\documentclass[10pt,letterpaper]{article}
\usepackage[top=0.85in,left=2.75in,footskip=0.75in]{geometry}

\usepackage{amsmath,amssymb}
\usepackage{ulem}

\usepackage{changepage}

\usepackage[utf8x]{inputenc}

\usepackage{textcomp,marvosym}

\usepackage{cite}

\usepackage{nameref,hyperref}

\usepackage[right]{lineno}

\usepackage{microtype}
\DisableLigatures[f]{encoding = *, family = * }

\usepackage[table]{xcolor}

\usepackage{array}

\newcolumntype{+}{!{\vrule width 2pt}}

\newlength\savedwidth

\usepackage[aboveskip=1pt,labelfont=bf,labelsep=period,justification=raggedright,singlelinecheck=off]{caption}

\makeatletter
\renewcommand{\@biblabel}[1]{\quad#1.}
\makeatother

\usepackage{lastpage,fancyhdr,graphicx}
\usepackage{epstopdf}
\fancyheadoffset[L]{2.25in}
\fancyfootoffset[L]{2.25in}
\begin{document}
\vspace*{0.2in}

\begin{flushleft}
{\Large
\textbf\newline{Past production constrains current energy demands: persistent scaling in global energy consumption and implications for climate change mitigation } 
}
\newline
\\
Timothy J. Garrett\textsuperscript{1},
Matheus Grasselli \textsuperscript{2},
Stephen Keen\textsuperscript{3},
\\
\bigskip
\textbf{1} Department of Atmospheric Sciences, University of Utah, Salt Lake City, UT, USA
\\
\textbf{2} Department of Mathematics, McMaster University, Hamilton, Ontario, Canada
\\
\textbf{3} Institute for Strategy, Resilience and Security, University College London, London,UK
\\
\bigskip

%
%





* tim.garrett@utah.edu

\end{flushleft}
\section*{Abstract}
Climate change has become intertwined with the global economy. Here, we
describe the importance of inertia to continued growth in energy consumption. Drawing from thermodynamic arguments, and using 38 years of available statistics between 1980 to 2017, we find a persistent time-independent scaling between the historical time integral $W$ of world inflation-adjusted economic production $Y$, or $W\left(t\right) = \int_0^t Y\left(t'\right)dt'$, and current rates of world primary energy consumption $\mathcal E$, such that  $\lambda = \mathcal{E}/W = 5.9\pm0.1$ Gigawatts per trillion 2010 US dollars. This empirical result implies that population expansion is a symptom rather than a cause of the current exponential rise in $\mathcal E$ and carbon dioxide emissions $C$, and that it is past innovation of economic production efficiency $Y/\mathcal{E}$ that has been the primary driver of growth, at predicted rates that agree well with data. Options for stabilizing $C$ are then limited to rapid decarbonization of $\mathcal E$ through sustained implementation of over one Gigawatt of renewable or nuclear power capacity per day. Alternatively, assuming continued reliance on fossil fuels, civilization could shift to a steady-state economy that devotes economic production exclusively to maintenance rather than expansion. If this were instituted immediately, continual energy consumption would still be required, so atmospheric carbon dioxide concentrations would not balance natural sinks until concentrations exceeded 500 ppmv, and double pre-industrial levels if the steady-state was attained by 2030. 



\section{Thermodynamic overview of civilization growth}

Like other biological systems \cite{Gowdy2013}, the human economy
interacts with its surroundings through flows of energy and matter
\cite{Job2006,Smith2008}. Collectively, we mine primary energy resources,
and use the energy to power civilization and convert raw materials into the material make-up of civilization or Earth's ``technosphere''
\cite{haff2014technology}. The circulations of our lives include
the back-and-forth material exchange of people, goods, and information
along transportation and communication networks, and our cardiovascular,
pulmonary and nervous systems \cite{buzsaki2004,barabasi1999emergence,Dijk2012,Johnson2012}.
These require a power source, and any impulse of energy that passes through these networks is ultimately dissipated through
frictional losses as waste heat. Concurrently, our infrastructure, bodies, and even our memories undergoes decay. So, without
a continual drawdown of primary energy and raw material resources
to continually rebuild civilization, it would inevitably fall apart.

\begin{figure}[ht]
\includegraphics[width=5in]{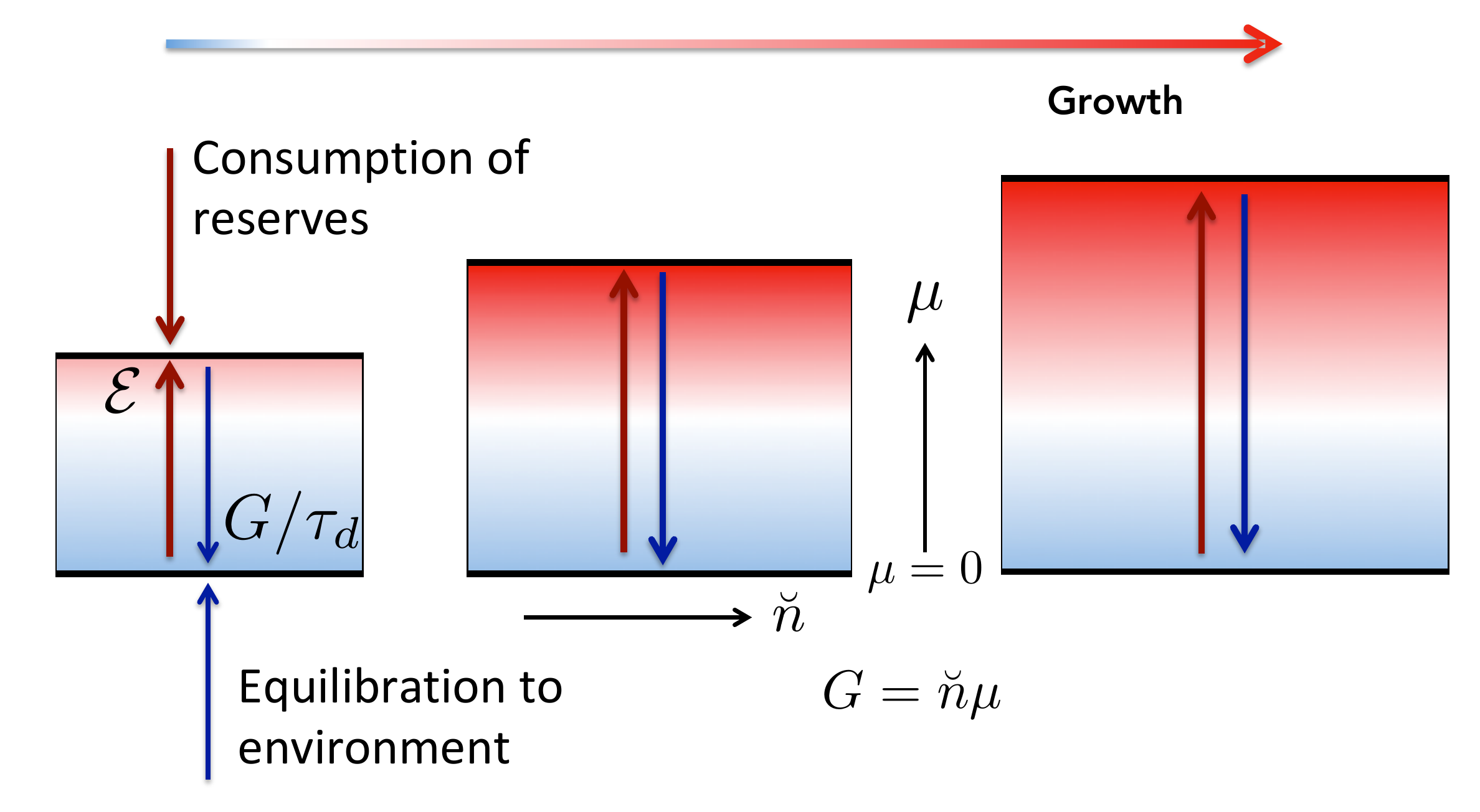}

\caption{Illustration of the short-term thermodynamic balance between primary
energy consumption $\mathcal{\mathcal{\mathcal{\mathcal{\mathcal{\mathcal{E}}}}}}$
and dissipation at rate $G/\tau_{d}$ (Eq. \ref{eq:dDGdt}) in a civilization
experiencing long-term, proportionate material $\breve{n}$ and specific
potential $\mu$ growth (Eq. \ref{eq:dDGdtlong}).\label{fig:Thermo}}
\end{figure}
While the entirety of humanity seems infinitely complex in its range of
activities, what all social phenomena have in common is their need for energy. The sum total of their power consumption is constrained by the world total, or the rate of primary energy consumption $\mathcal {E}$. By consuming primary energy, for example through combustion, civilization is stretched to a potential $G$, higher than an equilibrium with the environment characterized by $G=0$ (Fig. \ref{fig:Thermo}). 

This potential energy $G$ is converted into civilization circulations
by doing reversible work $\mathcal{W}^{rev}$ to move civilization
elements along pathways that align with the potential gradient $\nabla G$
\cite{AnnilaSalthe2009} and, through frictional losses, work is
converted to waste heat that is radiated to space. If the average
velocity of the circulations is $\vec{v}$ , and no primary energy
is consumed, then the rate at which potential energy is   dissipated
is
\begin{equation}
\mathcal{W}^{rev}=-\frac{dG}{dt}=
\vec{v}\cdot\nabla G= \frac{G}{\tau_{d}}\label{eq:reversible work}
\end{equation}
where the dissipation time is
$\tau_{d}=1/\left(\vec{v}\cdot\nabla\ln G\right)$.

The reason civilization does not relax to an environmental equilibrium
is that sustained primary energy consumption continually elevates $G$. For example,
oil is extracted from a well for consumption, and there is back-pressure
in the reserves to replenish supplies. Available statistics \cite{AER2011}
suggest that civilization has been consuming energy nearly as fast
as it has been produced. Since 1980, the mean ratio of global production
to consumption has been 0.998 with an annual standard deviation smaller than 1\%.
So, while $G$ may fluctuate because consumption and dissipation are
out of phase over rapid timescales shorter than $\tau_{d}$, averaged
over timescales a bit longer, say one week, these variations are no
longer apparent. Then, the quasi-equilibrium condition for civilization
sustenance is:
\begin{equation}
\left(\frac{dG}{dt}\right)^{sust}=\mathcal{E}^{sust}-\frac{G}{\tau_{d}}\simeq0\label{eq:dDGdt}
\end{equation}
implying that $\mathcal{E}^{sust}=G/\tau_{d}$ is the required energy
consumption to sustain civilization circulations, or in terms of the
thermodynamic First Law, to maintain a short-term balance between
external heating and doing reversible work $\mathcal{W}^{rev}$.

Civilization has a wide range of activities, each with their own energetic
demands and timescales that could be characterized by a power spectrum of consumption $\mathcal{\mathcal{E}}_{\tau}^{sust}$
satisfying 
\begin{equation}
\mathcal{E}^{sust}=\int_{0}^{\infty}\mathcal{E}_{\tau}^{sust}\tau d\tau
\end{equation}
in which case the characteristic timescale for civilization is $\tau_{d}=\int_{0}^{\infty}\mathcal{E}_{\tau}^{sust}\tau d\tau/\mathcal{E}^{sust}$.
Given that approximately 40\% of our time is spent alternating between
our top two preferred locations, (e.g. home and work) \cite{Gonzalez2008},
$\tau_{d}$ might be taken to be about 1 day. Currently, the world
primary energy consumption rate is roughly 20 TW, so it follows from
the quasi-equilibrium relationship 
Eq. (\ref{eq:dDGdt}) that \begin{equation}
    G = \mathcal{E}^{sust}\tau_d \simeq 20\times10^{12}\:{\rm J\:s^{-1}}\times86400\:{\rm s}=1.7\times10^{18} \:{\rm J},
\end{equation}
 equivalent to the energy contained in 270 million barrels of oil.

Obviously, such high potential is a far cry from where we started in the
Stone Age. How did we become so strong? There has been no external
hand to turn up the civilization flame. Purely mathematically,
reaching our current state meant accumulating successive increments
in potential $G$ over civilization's history: 
\begin{equation}
G\left(t\right)=\int_{0}^{t}\frac{dG\left(t'\right)}{dt'}dt'\label{eq:DG}
\end{equation}
Eq. \ref{eq:dDGdt} implies that the primary energy supply $\mathcal{E}$ must have
been greater than humanity's collective metabolic needs  $\mathcal{E}^{sust}$ so that
\begin{align}
\frac{dG}{dt} & =\frac{G}{\tau_{long}}=\mathcal{E}-\mathcal{E}^{sust}>0\label{eq:dGdt total}
\end{align}
An important point here is that civilization growth timescales $\tau_{long}$
are decades to centuries, that is, much longer than $\tau_{d}$. For
example, recent growth rates of global primary energy consumption
are approximately 2\% per year (or equivalent to $\tau_{long}\simeq 50$ years), so the
implied difference between $\mathcal{E}$ and $\mathcal{E}^{sust}$
is only about 0.01\%. Such growth might be imperceptible in our daily lives,
but it does slowly accumulate. We can therefore write 
\begin{equation}
    \mathcal{E}^{sust}=(1-\epsilon)\mathcal{E},
\end{equation}
where $\epsilon=\tau_{d}/(\tau_d+\tau_{long})\ll1$, so that the approximation $\mathcal{E}\simeq\mathcal{E}^{sust}$ can be made that at any given time. For the remainder of $\mathcal E$, we have
\begin{equation}
\mathcal{W}^{irr}= \frac{dG}{dt} = \epsilon\mathcal{E}, \label{eq:dGdt epsilon}
\end{equation}
so that $\epsilon$ can be seen to be the efficiency of converting
primary energy consumption $\mathcal{E}$ to the irreversible work $\mathcal{W}^{irr}$
that enables growth. Note how the sign on $\mathcal{W}^{irr}$ is opposite
to that in Eq. \ref{eq:reversible work}, where potential energy is
dissipated to do the reversible work $\mathcal{W}^{rev}$ that maintains civilization
circulations. Instead, irreversible work is done to grow the civilization potential $G$. Moreover, it follows that the growth rate of the potential is given by 
\begin{equation}
    \eta_G := \frac{d\ln G}{dt}=\frac{1}{G}\frac{dG}{dt}= \frac{1}{\tau_{long}} =\frac{\epsilon}{(1-\epsilon)\tau_d}\simeq \frac{\epsilon}{\tau_d}
\end{equation}
Similarly, taking the derivative of
\begin{equation}
\mathcal{E} =\frac{\mathcal{E}^{sust}}{1-\epsilon}=
\frac{G}{(1-\epsilon)\tau_{d}}\label{eq:E nu}
\end{equation}
we find that the growth rate of primary energy consumption is 
\begin{equation}
\eta_{\mathcal{E}} := \frac{d\ln\mathcal{E}}{dt}=\frac{1}{\mathcal{E}}\frac{d\mathcal{E}}{dt}=\frac{\epsilon}{(1-\epsilon)\tau_{d}}\left(1+\tau_d \eta_\epsilon\right) \simeq \frac{\epsilon}{\tau_d}
\label{eq:dEdt}
\end{equation}
where 
\begin{equation}
    \eta_\epsilon := \frac{d\ln \epsilon}{dt}
\end{equation}
is the growth rate of $\epsilon$ and $\tau_d \eta_{\epsilon} \ll 1$.

The mechanism through which an imbalance leading to $\mathcal{W}^{irr}>0$ should emerge is not discussed
in detail here. It follows from discovering accessible energy resources faster than
they are consumed \cite{GarrettEF2014, GarrettHindcasts2015}. The gradient $\nabla G$ becomes steeper nearer the
energy source than to the dissipative sink, so that there is a net convergence of energy in civilization. Very generally, this process is analogous to the heat equation $\left(dG/dt\right)_{long}=\nabla\cdot\left(\mathcal{D}\nabla G\right)=\mathcal{D}\nabla^{2}G$,
where $\mathcal{D}$ is a constant diffusivity. 

For more intuitive insight, consider a child as a more familiar complex system. 
From Eq. \ref{eq:dDGdt}, a child with potential
$G$ has quasi-equilibrium metabolic needs of $\mathcal{E}^{sust}=G/\tau_{d}$
-- say about 50 Watts, or about 500
kJ per day per kg. Any food energy the child consumes is
used to do reversible work $\mathcal{W}^{rev}$ to maintain rapid
internal neurological, respiratory, cardiovascular, and nervous circulations
and to reconstitute food nutrients into its material makeup. If the
energy in the proteins, carbohydrates and fats of accessible food
is only just sufficient to offset decay and keep the child alive, then the child is at steady-state and $G$ and
$\mathcal{\mathcal{E}}^{sust}$ do not change. A healthy child, however,
does net irreversible work at rate $\mathcal{W}^{irr}=\epsilon \mathcal{E}$
to convert some small portion of the energy in food into accumulation
of body matter through a conversion factor of about 30 MJ kg$^{-1}$. The child eventually becomes a robust adult with higher
daily energy demands and the growth rate (hopefully) stabilizes. But even then, a typical rate of weight gain translates to a value of ${W}^{irr}/W^{rev}$ of about 0.2\%, while seemingly tiny, results in a 10 kg gain in mass, or 300 MJ of energy, over the 50 years span of a typical adult life \cite{Frayn2009}. 

This treatment suggests how to consider the coupling of energy and matter in an open system such as civilization. $G$ is a total potential,
so it can be decomposed into any arbitrary number of sub-components
with $G=\sum\breve{n}_{i}\mu_{i}$, depending on how closely civilization
is resolved. Here we take the simplest possible approach which is to
suppose, as illustrated in Fig. \ref{fig:Thermo}, that the accessibility
of energy by civilization can be defined by an interface with resources composed
of $\breve{n}$ material elements each with average potential $\mu$, so that 
\begin{equation}
    G(t) = \breve{n}(t)\mu(t).
    \label{eq:G_n_mu}
\end{equation}
Thus, civilization elements are not a purely
additive summation of civilization ``things'' $n$. Rather they
represent a number of network nodes, defined in terms of people, firms,
or nations that collectively do work to dissipate energy at rate
$\mathcal{E}^{sust}$. 

It therefore follows from \ref{eq:dGdt epsilon} and \ref{eq:G_n_mu} that any long-term net convergence $\epsilon\mathcal{E}$
is a surplus that can be partitioned between manufacturing more civilization
nodes or increasing their average potential $\mu$: 
\begin{equation}
\mathcal{\epsilon\mathcal{E}} =\mathcal{W}^{irr}=\frac{dG}{dt}=\mu\frac{d\breve{n}}{dt}+\breve{n}\frac{d\mu}{dt}\label{eq:dDGdtlong}
\end{equation}
By increasing the average potential at rate $d\mu/dt$, existing civilization
elements $\breve{n}$ can go farther and faster. On the other hand, production of new civilization elements $d\breve{n}/dt$ with average potential $\mu$ implies a phase change. Just as an excess 30 MJ 
of energy is required for the chemical transformation of food into
a kilogram of flesh, stationary raw materials such as forests, fish stocks,
and iron ore are rearranged into familiar forms such as cars, roads,
and communications systems. 

We assume the following thermodynamic proportionality:
\begin{equation}
\frac{d\mu}{d\breve{n}}=(k-1)\frac{\mu}{\breve{n}}\label{eq:proportionality}
\end{equation}
for a constant $k$. Substituting this into Eq. \ref{eq:dDGdtlong} leads to 
\begin{equation}
    \mathcal{W}^{irr}=k\mu\frac{d\breve{n}}{dt}
\end{equation}
or rearranging, net production of network nodes is related to current
energy consumption through
\begin{align}
\frac{d\breve{n}}{dt} & =\frac{\epsilon}{k\mu}\mathcal{E}\label{eq:dndt}
\end{align}
where $\epsilon/\left(k\mu\right)$ is the efficiency of converting primary energy consumption to network growth.

To summarize, we consume energy at rate 
$\mathcal{E}$. Most of it, namely $\mathcal{E}^{sust}=(1-\epsilon)\mathcal{E}$, is used to do reversible
work at rate $\mathcal{W}^{rev}$ to sustain circulations along civilization
networks. The surplus $\epsilon\mathcal{E}$ is used to do irreversible work
at rate $\mathcal{W}^{irr}$ to convert raw materials into the
stuff of humanity and extend existing network nodes at rate $d\breve{n}/dt$.
These form the fabric of society, including roads, telecommunications
networks and even neural pathways encapsulating memories within our
brains. We grow fastest if the efficiency $\epsilon$ is high. And
through expansion of the physical interface at rate $d\breve{n}/dt$
with reserves of energy and matter, civilization grows, leading recursively
to further expansion and higher consumption. 

\section{Thermodynamics of global economic value}

Can these strictly thermodynamic concepts be linked to the economy
expressible in financial terms? Suppose for the moment that they can,
and that there exists a hypothetical global quantity expressible in
units of real currency (or ``widgets'') $W$ that expresses the size
of civilization and is presumed to be proportional to energy consumption
and the civilization potential through a constant scaling factor $\lambda$:
\begin{equation}
\frac{G}{\tau_{d}}\simeq \mathcal{E}=\lambda W
\label{eq:E lambda W}
\end{equation}
In this expression, civilization elements of whatever kind contribute
to $W$ only insofar as they contribute to the overall network capacity
to dissipate primary potential energy at rate $\mathcal{E}$, requiring
a thermodynamic potential $G$.

From a purely dimensional perspective, the simplest possible economically
quantifiable definition of $W$ is that it is an integration over
time of a global quantity with units of widgets per time. The approach
previously taken in \cite{GarrettCO2_2009} was to suppose that the
most obvious candidate is the world economic production $Y$ (or gross
domestic product GDP) calculated at market exchange rates (MER) and
adjusted for inflation, in which case $W$ is the world cumulative
production:
\begin{equation}
W=\int_{0}^{t}Y\left(t'\right)dt'
\label{eq:WintY}
\end{equation}
From  Eqs. \ref{eq:E lambda W} and \ref{eq:WintY}, the testable hypothesis
is that economic production aggregated for all nations and integrated
over all of history is tied through a constant $\lambda$ to energy
consumption through: 
\begin{equation}
\mathcal{E}\left(t\right)=\lambda\int_{0}^{t}Y\left(t'\right)dt'\label{eq:EintY}
\end{equation}
Note the similarity here with Eq. \ref{eq:DG}. If the expression in Eq. \ref{eq:EintY}
can be shown to empirically justified, it would imply that there is
a constant relationship between inflation-adjusted economic production
$Y$ and a \emph{rise }in global energy consumption demands $\mathcal{E}$:
\begin{equation}
Y=\frac{dW}{dt}=\frac{1}{\lambda}\frac{d\mathcal{E}}{dt}\label{eq:YdEdt}
\end{equation}
From Eqs. \ref{eq:dEdt} and \ref{eq:dndt}, we can see that economic production is
tied to the use of primary energy to do irreversible work to
convert raw materials into civilization matter. Namely, 
\begin{equation}
Y=\frac{\mathcal{E}\eta_{\mathcal{E}}}{\lambda}\simeq
\frac{\epsilon\mathcal{E}}{\lambda\tau_d}=
\frac{k\mu}{\lambda\tau_{d}}\frac{d\breve{n}}{dt}=\frac{\mathcal{W}^{irr}}{\lambda\tau_{d}}\label{eq:Y_J}
\end{equation}
In other words, with a surplus of energy $\epsilon\mathcal{E}$, the extraction of
raw materials can be tied to the production of economically useful material goods (namely  ``widgets'') as has been noted previously in \cite{Wiedmann2015}.
These add value by speeding up human activity and increasing civilization
size. Economic production provides the recipe for growth by expanding
our collective interface with energy and material reserves, leading
to positive increments in our capacity to consume.

\begin{figure}[ht]
\includegraphics[width=12cm]{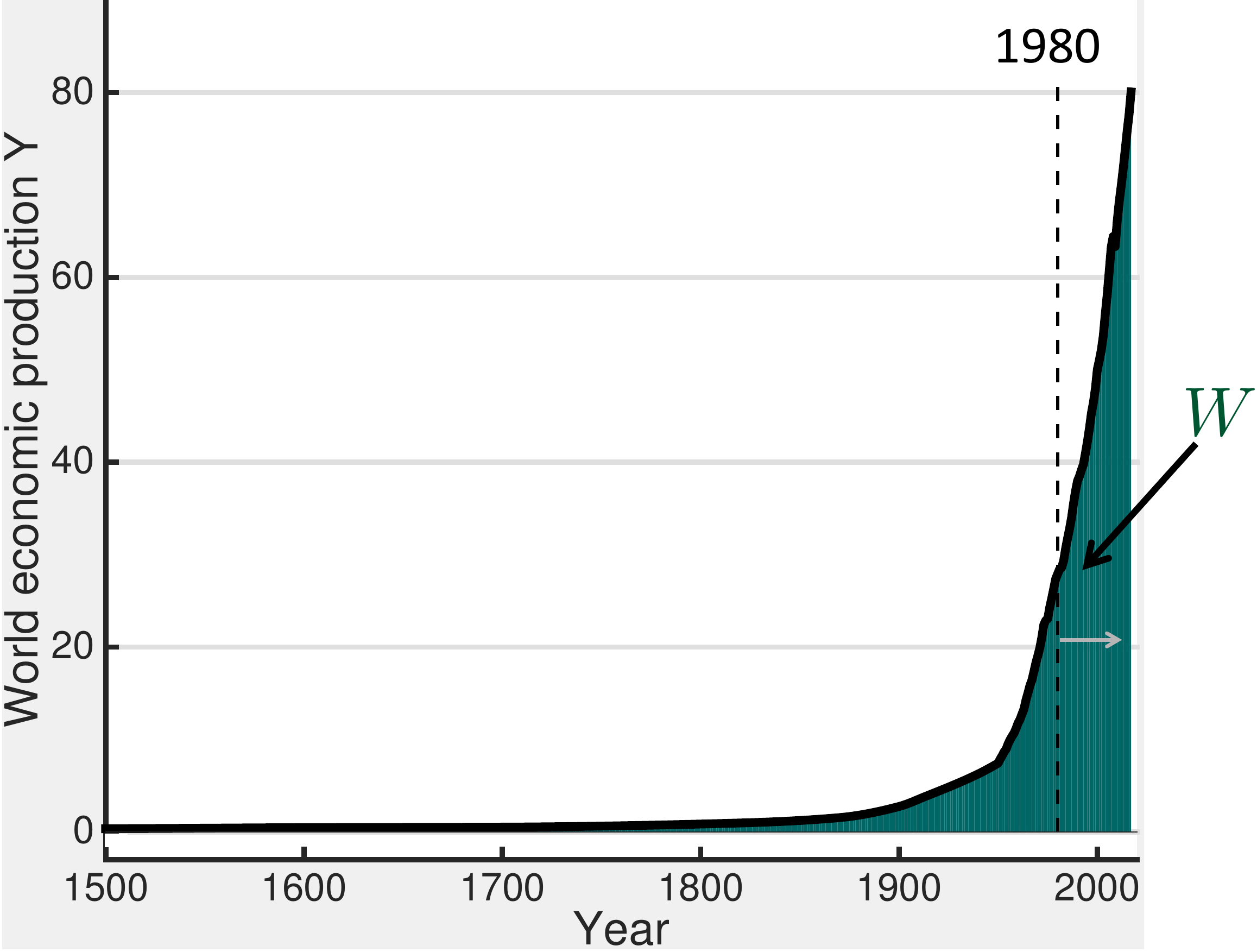}

\caption{Evolution of world economic production $Y$ in trillion 2010 USD per
year (solid line) and the integrated contribution to the world cumulative
production $W$ in trillion 2010 USD proportional to the shaded area under the curve. The period between 1980 to 2017 that is used
for comparison with world primary energy consumption $\mathcal{E}$
as described in the text is delineated by the dashed line and shown by the gray arrow. \label{fig:World-economic-production}}
\end{figure}

Available data for testing Eq. \ref{eq:EintY} include annual market
exchange rate estimates of GDP, inflation-adjusted to ``real'' units (namely constant 2010 US dollars) from the World Bank \cite{WorldBank2019} and the
United Nations \cite{UNstats} for the years 1970 to 2017,
and reconstructions from the Maddison database of the real GDP adjusted
for purchasing power parity (PPP) 1990 USD for each year between 1950
and 1992, with more sparse estimates extending back to 1 CE \cite{Maddison2003} (see details in Appendix \ref{S1_Appendix}). Annual rates of global primary energy consumption
$\mathcal{E}$ are available from British Petroleum
and the U.S. Department of Energy (DOE) Energy Information Administration
(EIA) for the time period 1980 to 2017 \cite{AER2011,BritishPetroleum2018}.
For initialization of the integration in Eq. \ref{eq:WintY}, it is
estimated that world cumulative production in 1 C.E.
was 250 trillion 2010 USD, a number that is obtained iteratively so
that there is consistency for that period between growth of $W$ and
population growth rates of about 6\% per century \cite{USCensus}.
This reconstruction is about 7.3\% of the value obtained for 2017,
suggesting the ancient world had already evolved non-negligible wealth
in its Western, Middle Eastern, and Eastern empires \cite{Marchetti2012}.
The Maddison database is sparse, and presumably increasingly uncertain
the farther one goes back in time. However, the value of $W$ accumulated
over the period 1 CE to 1000 CE is just 4.6\% of the value in the
year 2017. It is not until the last century that $W\left(t\right)$
grows appreciably (Figure \ref{fig:World-economic-production}). The
world cumulative production between 1980 and 2017 comprises
a remarkable 60\% of the historically accumulated total. 
\begin{figure}[ht]
\includegraphics[width=12cm]{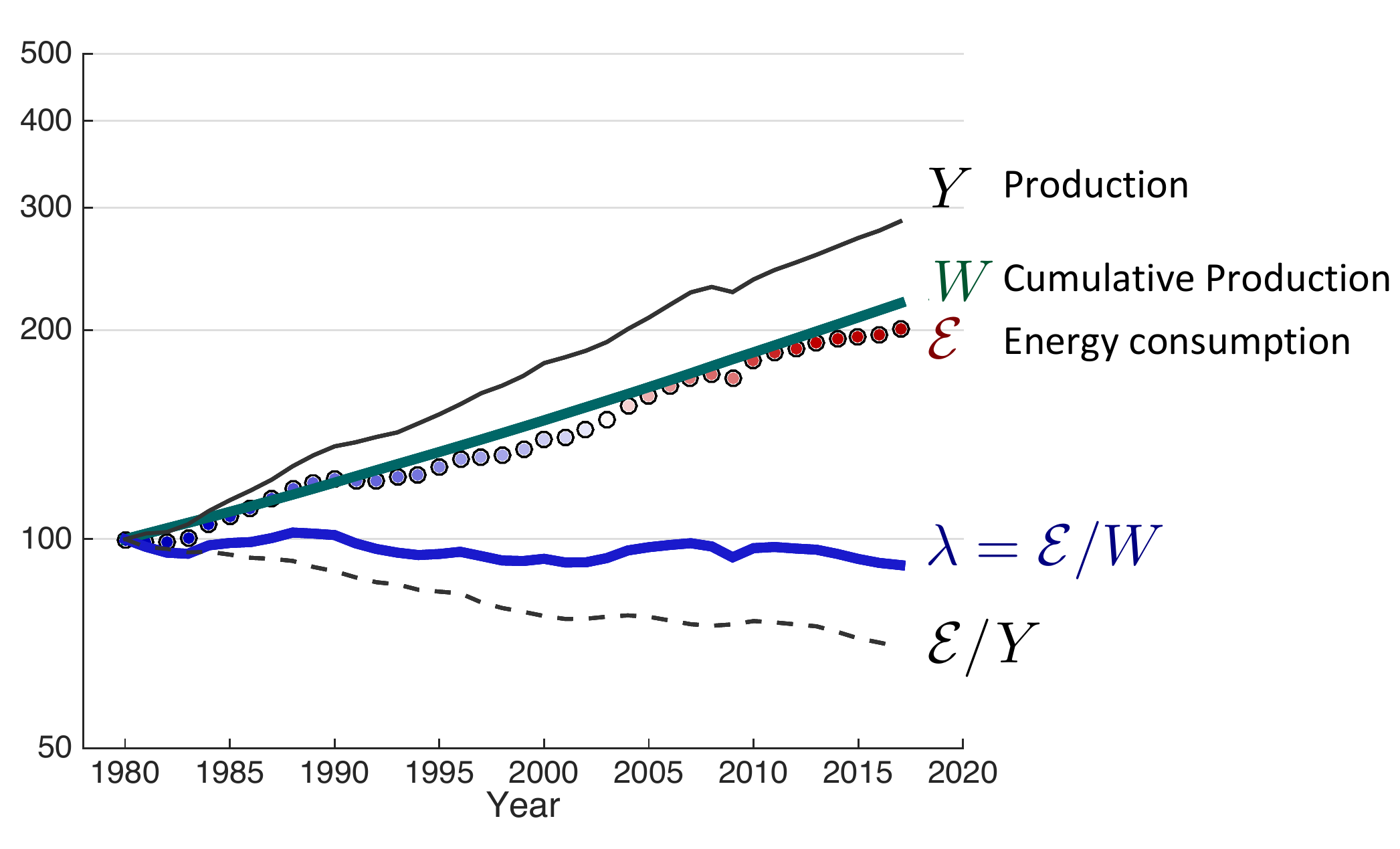}

\caption{Relative evolution since 1980 of the world real GDP $Y$, economic
potential $W=\int_{0}^{t}Y\left(t'\right)dt'$ (Eq. \ref{eq:WintY}),
primary energy consumption $\mathcal{E}$, $\lambda=\mathcal{E}/W$
(Eq. \ref{eq:E lambda W}) and the energy intensity of production
$\mathcal{E}/Y$.\label{fig:Relative-evolution}}
\end{figure}
\begin{table}[ht]
\centering
\small
\caption{\label{tab:lambda}The global value of $\lambda$ ((gigawatts per
trillion 2010 US dollar with standard deviation) defined by Eq. \ref{eq:E lambda W}
for various time periods.}
\begin{tabular}{c|c|c|c|c|c|c}
\hline 
Period & 1980-1990 & 1990-2000 & 2000-2010 & 2010-2017 & 1980-2010 & 1980-2017\tabularnewline
\hline 
$\lambda=\mathcal{E}/W$ & 6.04$\pm0.14$ & 5.83$\pm$0.15 & 5.83$\pm$0.14 & 5.79$\pm$0.14 & 5.90$\pm0.16$ & 5.88$\pm0.17$\tabularnewline
\hline 
\end{tabular}
\end{table}

The relative evolution since 1980 of $\mathcal{E}$, $Y$, $W$, $\lambda=\mathcal{E}/W$
and $\mathcal{E}/Y$ is shown in Figure \ref{fig:Relative-evolution}
and summarized in Table \ref{tab:lambda}. Expressing  Eq. \ref{eq:EintY}
as a summation of yearly data, $\lambda_{j}=\mathcal{E}_{j}/\sum_{0}^{j}Y_{i}$,
the mean value\footnote{All values for $\lambda$ in this paragraph have units of gigawatts per trillion 2010 USD} of $\lambda$ is 5.9 with a standard deviation of 0.1 (2\%) and an uncertainty in the mean at the 95\% confidence
level of 0.05 (0.8\%). As an
indicator of the sensitivity to uncertainty in the data, assuming for the initializing of the integration a value of $W\left(1\right)$ double the previously derived value
of 250 trillion 2010 USD, then $\lambda=5.2\pm0.2$. If half as much, then $\lambda=6.2\pm0.2$. While there is arguably a
small secular downward trend of 0.1\% yr$^{-1}$, the temporal variation
in $\lambda$ is sufficiently small and insensitive to assumptions
that it appears useful as a scaling factor relating an economic quantity
$W$ to a physical measure $\mathcal{E}$. If $\tau_{d}\simeq1$ day,
then the implication is that $G/W=\lambda\tau_{d}\simeq510\pm9$ Joules
per 2010 US dollar.

The relationship in Eq. \ref{eq:EintY} appears to hold empirically. But it might seem surprising given that it lacks any explicit representation of the spontaneous appreciation or of consumption,
decay, and depreciation. Obviously, trade agreements and resource
discovery add value without being tabulated in production. And not
all of what has been produced remains as old technologies turn obsolete
and networks fray due to physical destruction by storms, rust, and
forgetting. A cook may produce food in a restaurant, but other
than temporary body sustenance and lingering memories, the meal is
gone. The total ``wealth'' of civilization cannot be simply a summation
of production. Equation \ref{eq:EintY} is easier to justify, however, if there is exists a steady fraction of production that contributes to civilization expansion through irreversible work, as in Eq. \ref{eq:Y_J}.

In traditional macroeconomic growth models, there is also a quantity
with units of widgets, termed real capital $K$ that relates to real economic output $Y$ according to some production function $Y=f\left(K,L\right)$
where $L$ represent labor. A common example is the Cobb-Douglas production
function  $Y=AK^{\alpha}L^{1-\alpha}$ where $\alpha$ is determined
from a fit to data and $A$ is termed a ``total factor productivity''
that can itself evolve with time according to, for example, investments
in research and development \cite{Romer1994} (see also \cite{Keen2019} for a different interpretation of $A$ in the context of a Cobb-Douglas production function taking into account energy inputs). One could argue that the appropriate theoretical relationship to draw is not between $\mathcal{E}$ and $W$ as defined in Eq. \ref{eq:WintY}, but between $\mathcal{E}$ and capital $K$. 

To explore this possibility further, observe that capital $K$ evolves as 
\begin{equation}
    \frac{dK}{dt}=I-\delta K = Y-\mathcal{C} -\delta K
    \label{eq:capital}
\end{equation} 
where $I=Y-\mathcal{C}$ is gross investment, $\delta$ is a depreciation rate, and $\mathcal{C}$ refers to consumption of goods that are assumed to depreciate much faster than capital itself. If we define $\gamma$ as the fraction of production used for consumption plus depreciation, that is 
\begin{equation}
    \gamma Y = \mathcal{C}+\delta K,
\end{equation}
then Eq. \ref{eq:capital} becomes $dK/dt = (1-\gamma) Y$, or equivalently,
\begin{equation}
    K\left(t\right) = \int_0^t \left(1 - \gamma\left(t'\right)\right) Y\left(t'\right)dt'.
\end{equation}
Provided $\gamma$ does not vary greatly in time, we can approximate the factor $(1-\gamma)$ by its time average $\left<1-\gamma\right>$ so that 
\begin{equation}
    K \simeq \left<1 - \gamma\right>\int_0^t Y\left(t'\right)dt'.
\end{equation}
In other words, if there were a fundamental scaling between $\mathcal{E}$ and $K$, this would also imply the scaling expressed in Eq.  \ref{eq:EintY}. 

Then, if society experienced a sudden collapse in its growth toward a steady-state economy with constant $\mathcal{E}$, economic production $Y$ could remain positive, even adjusting for inflation, but the implication would be that all production would be used simply for consumption and to sustain capital $K$ at its current state, that is ot say, $\gamma\left(t\right) = 1$, and both $K$ and $\mathcal{E}$ are stationary. 

Unfortunately long-term global time series to test whether $K$ scales with $\mathcal{E}$ are scant, in part because there is   disagreement among economists about how to appropriately aggregate the value of items as different as houses and   tractors \cite{Harcourt1972},   and also because of difficulties with estimating a starting point for the corresponding time series \cite{LeonLedesmaMcAdamWillman2010} similarly to the difficulty we mentioned above related to $W(1)$. Also, it is unclear what to include because economic capital does not normally consider for example people or their culture, even though these are core elements of the human dissipative engine. In other words, whereas capital $K$ and the global quantity $W$ in Eq. \ref{eq:E lambda W} may be numerically related, they are not conceptually analogous. The value expressed by $W$ lies not
so much in an aggregation of inert ``things'', but in their summed capacity to form interconnected networks through a thermodynamic potential $G=\breve{n}\mu$ that sustains the dissipation of potential energy at rate $\mathcal{E}$, whether it is from international trade, housework, or the firing of neuronal networks in human brains. 

In the meantime, even if $W$ calculated from Eq. \ref{eq:WintY} yields an overestimate of the true aggregated wealth of civilization, $\lambda$ as
calculated in Eq. \ref{eq:EintY} has been nearly a constant for sufficiently long that it seems reasonable to assume that this regularity will carry into the future.

\section{Implications for economic growth}

Adopting a nearly fixed relationship between $W$ and $\mathcal{E}$ offers
a testable basis for exploring thermodynamic constraints on economic
growth. From Eqs. \ref{eq:WintY} and \ref{eq:EintY}, both the
world cumulative production and primary energy consumption grow at
the exponential rate:
\begin{equation}
\eta_{\mathcal{E}}=\eta_{W}:=\frac{1}{W}\frac{dW}{dt} = \frac{Y}{W}=\lambda\frac{Y}{\mathcal{E}}= \lambda\varepsilon \label{eq:eta Y ratio}
\end{equation}
where $\varepsilon:=Y/\mathcal{E}$ is the energy efficiency of economic production or energy productivity. That is to say, higher energy productivity is related with increased energy demands. 

This result may seem counter-intuitive, but it rests only
on the empirical result that $\lambda$ is a constant. It should not
be confused with the concept of backfire or Jevons' Paradox \cite{Jevonsbook2007,Jenkins2011,Sorrell2014},
which also argues that efficiency gains lead to consumption growth,
but is stated within the context of traditional economic models for
specific economic sectors or nations and is often refuted. Here, Eq.
\ref{eq:eta Y ratio} applies strictly within a global context
so that complications due to trade do not play a role.
Interpreted physically, Eq. \ref{eq:eta Y ratio} can be compared
with Eq. \ref{eq:Y_J}, suggesting that $\varepsilon \simeq \epsilon/\lambda\tau_{d}$, or that energy productivity is related to the efficiency of doing irreversible work $\mathcal{W}^{irr}=\mathcal{\epsilon\mathcal{E}}$ that further expands the capacity to consume. The implication is that civilization is an emergent phenomenon that
grew spontaneously to its currently high state of production through
an energy surplus. 

The decades following 1950 -- known as the ``great
acceleration'' -- stand out in particular, when a relative ease
of access to oil led to rapid innovations \cite{Steffen2015,GarrettHindcasts2015}.
To characterize acceleration in growth, $\eta_{W}$ can be considered
to have its own exponential growth rate 
\begin{equation}
\eta_{I}:=\frac{d\ln\eta_{W}}{dt} \label{eq:Innovationrate}
\end{equation}
We term this an ``innovation rate'' because From Eq. \ref{eq:eta Y ratio},
if $\lambda$ is a constant, $\eta_{I}$ can also be expressed as
\begin{equation}
\eta_{I}=\eta_\varepsilon := \frac{d\ln\varepsilon}{dt} \label{eq:innovation efficiency}
\end{equation}
Whenever conditions support an increase in energy productivity $\varepsilon$
(or thermodynamic efficiency $\epsilon$), for example due to energy
reserve discoveries \cite{GarrettHindcasts2015}, then $\eta_{I}>0$
and civilization growth is superexponential.

The world GDP growth rate $\eta_{Y}:=d\ln Y/dt$ follows a slightly
different pathway. From $Y=\varepsilon\mathcal{E}$ and Eq. \ref{eq:eta Y ratio}, it follows that:
\begin{equation}
\eta_{Y}=\eta_\mathcal{E} + \eta_{\varepsilon} = \lambda\varepsilon + \eta_{\varepsilon} 
\label{eq:GDP growth efficiency}
\end{equation}
At global levels, by reducing the energy required for manufacture
of valuable goods, civilization bootstraps through innovation to higher
GDP growth and accelerating rates of energy consumption.

\begin{table}[ht]
\centering
\caption{\label{tab:Rates} Measured average growth rates (\%/yr) compared with rates derived assuming
$\lambda$ is a constant in bold. Pertinent equations are
in parentheses.}

\begin{tabular}{c|c|cc|cc|cc}
\hline 
 & Wealth & \multicolumn{2}{c|}{Energy} & \multicolumn{2}{c|}{Innovation} & \multicolumn{2}{c}{GDP}\tabularnewline
\hline 
 & $\eta_{W}$  & $\eta_{\mathcal{E}}$ & $\lambda\varepsilon$ (\ref{eq:eta Y ratio}) & $\eta_{I}$ (\ref{eq:Innovationrate}) & \textbf{$\eta_{\varepsilon}$ }(\ref{eq:innovation efficiency}) & $\eta_{Y}$ &  \textbf{$\lambda\varepsilon+\eta_{\varepsilon}$ }(\ref{eq:GDP growth efficiency})\tabularnewline
1980-2010 & 2.06 & 1.98 & \textbf{2.09} & 0.82 & \textbf{0.91} & 2.88 & \textbf{3.0}\tabularnewline
2010-2017 & 2.33 & 1.60 & \textbf{2.40} & 0.45 & \textbf{1.18} & 2.78 & \textbf{3.58}\tabularnewline
1980-2017 & 2.14 & 1.84 & \textbf{2.15} & 0.73 & \textbf{1.0} & 2.84 & \textbf{3.15}\tabularnewline
\hline 
\end{tabular}
\end{table}

Table \ref{tab:Rates} shows a check of the applicability of the derived
growth equations \ref{eq:eta Y ratio} to \ref{eq:GDP growth efficiency}
for the period between 1980 and 2010 and a shorter more recent period
from 2010 to 2017. Overall, there is close agreement between observations
and calculated rates of change based on the constancy of $\lambda$
suggesting the relationship offers a useful tool for making simplified
predictions. However, there are also some discrepancies after 2010.
For example, as stated in Eq. \ref{eq:eta Y ratio}, a consequence
of a constant value for $\lambda$ is that the primary energy consumption
growth rate $\eta_{\mathcal{E}}$ should be equivalent to both the
growth rate of the world cumulative production $\eta_{W}$ and the product $\lambda\varepsilon$.
For the period 1980 to 2010 over which energy consumption increased
by 80\%, the three calculations differ by at most $6\%$. 

Similarly the GDP growth rate calculated   from the expression 
$\lambda\varepsilon + \eta_\varepsilon$ (rightmost in Eq. \ref{eq:GDP growth efficiency}) agrees to within 4\% with the
directly calculated value $\eta_Y$. However, for the shorter period between
2010 and 2017, the agreement is less precise, and also
for the time period 1980 to 2017. The
reason is unknown, although note from Table \ref{tab:Rates} and the mathematical equivalency $\eta_Y = \eta_\mathcal{E} + \eta_\varepsilon$ that any discrepancy between $\eta_Y$ and $\lambda\varepsilon + \eta_\varepsilon$ is due entirely to discrepancies between $\eta_\mathcal{E}$ and $\lambda\varepsilon $. Energy consumption increased by just 11\% between 2010 and 2017 so there is greater susceptibility to quantification errors. Also, nominal GDP is what is measured, and any calculation of real $Y$ requires an accurate assessment of the GDP deflator that attempts to account for inflation, and its true magnitude may have been underestimated. Indeed, $\lambda = \mathcal{E}/W$ was 2\% lower in the latter period (Table \ref{tab:lambda}), and any departure from constancy affects calculation of higher order derivatives such as the growth rates $\eta_{\mathcal{E}}$ and $\eta_Y$. 

\section{Implications for carbon dioxide emissions and concentrations}

\subsection{Simplifications to socio-economic drivers}

The anthropogenic contribution to atmospheric carbon dioxide emissions
can be conveniently decomposed into the product of population $P$,
affluence expressible as the gross domestic product (GDP) per person
$g:=Y/P$, the energy intensity of economic production $i=:\mathcal{E}/Y=1\varepsilon$,
and the amount of carbon dioxide emitted by the choice of energy source
$c:=C/\mathcal{E}$, leading to 
the Kaya identity $C=P\times g\times i\times c$ \cite{Raupach2007}.
In terms of  
growth rates, 
we have
\begin{equation}
\eta_{C}=\eta_{P}+\eta_{g}-\eta_{\varepsilon}+\eta_{c}\label{eq:Kaya eta}
\end{equation}
The most recent IPPC report \cite{IPCC_WG32014} lists rapid increases
in population and standard of living as primary drivers of past emissions
growth but focuses on innovations that improve production efficiency
and reductions in the carbon intensity of fuels as targets for future
reductions.

Eq. \ref{eq:Kaya eta} helps frame issues surrounding climate change
mitigation. But it is only a mathematical identity and, as such, it does not
directly allow for dynamic interactions between terms. What the link expressed in  
Eq. \ref{eq:EintY} between current consumption and the economy provides
is an added strong constraint on interrelationships between population,
standard of living, and production efficiency. For example, inserting Eq. \ref{eq:EintY} in $C=c\mathcal{E}$ yields that 
current carbon dioxide emissions can be related to \emph{past } accumulated economic production through 
\begin{equation}
C\left(t\right)=\lambda c\int_{0}^{t}Y\left(t'\right)dt'\label{eq:C int Y}
\end{equation}
A revised expression then follows for Eq. \ref{eq:Kaya eta}, namely  
\begin{equation}
\eta_{C}=\eta_{c}+\eta_{\mathcal{E}} = \eta_{c}+ \lambda\varepsilon
\label{eq:Kaya new}
\end{equation}
where we used Eq. \ref{eq:eta Y ratio}. 

A test of Eq. \ref{eq:Kaya new} is shown in Table \ref{tab:Rates-C},
based on the values for $\lambda\varepsilon$ shown
in Table \ref{tab:Rates} and data from the Global Carbon Atlas \cite{GlobalCarbonAtlas2020}.
Between 1980 and 2010, the observed average annual rate of emissions
growth $\eta_C$ was 1.77 \% yr$^{-1}$, within 6\% of the calculate value of 1.88 \%
yr$^{-1}$ for $\eta_{c}+ \lambda\varepsilon$. For more
recent years, the discrepancy is 20\%.

\begin{table}[ht]
\centering
\small
\caption{\label{tab:Rates-C}Average growth rates in carbonization and CO$_{2}$
emissions (\%/yr). Rates derived assuming $\lambda$ is a constant
are shown in bold, and pertinent equations in parentheses. The units of $C/W$ are Gt C yr$^{-1}$ per quadrillion 2010 USD.}

\begin{tabular}{cc|c|c|c}

\hline 
 & Scaling (Eq. \ref{eq:C int Y}) & Carbonization $c=C/\mathcal{E}$ & \multicolumn{2}{c}{CO$_{2}$ emissions $C$}\tabularnewline
\hline 
 &  $C/W=\lambda c$ ($\pm$std. dev.) & $\eta_{c}$ & $\eta_{C}$ & \textbf{$\eta_{c}+\lambda\varepsilon$ }(\ref{eq:Kaya new})\tabularnewline
1980-2010 & 1.50$\pm0.06$ & -0.21 & 1.77 & \textbf{1.88}\tabularnewline
2010-2017 & 1.45$\pm0.05$ & -0.36 & 1.25 & \textbf{2.04}\tabularnewline
1980-2017 & 1.49$\pm0.06$ & -0.25 & 1.59 & \textbf{1.90}\tabularnewline
\hline 
\end{tabular}
\end{table}

Similarly, the revised identity for CO$_{2}$ emissions given by Eq. \ref{eq:Kaya new}
implies that 
\begin{equation}
\eta_{P}+\eta_{g}=\lambda\varepsilon+\eta_{\varepsilon}\label{eq:etap and eta g}
\end{equation}
A test of this relation is shown in Table  \ref{tab:population}. For the 1980 to
2017 time period, the difference between both sides of the equality
in Eq. \ref{eq:etap and eta g} is about 10\%. Also, note that over
the longer term $\eta_{p}$ and $\eta_{g}$ are remarkably similar. If $\breve{n}$ can be related to population and $\mu$ to standard
of living, this empirical result is consistent with the thermodynamic relationship given by Eq. \ref{eq:proportionality} with $k=2$.

\begin{table}[ht]
\centering
\caption{\label{tab:population}Average growth rates of population and standard
of living (\%/yr). Summed rate derived assuming $\lambda$ is a constant
from efficiency estimates are shown in bold.}

\begin{tabular}{cc|c|c|c}
\hline 
 & Population  & Standard of living & \multicolumn{2}{c}{Summation}\tabularnewline
\hline 
 & $\eta_{P}$ & $\eta_{g}$ & $\eta_{p}+\eta_{g}$ & \textbf{$\lambda\varepsilon+\eta_{\varepsilon}$ }(\ref{eq:etap and eta g})\tabularnewline
1980-2010 & 1.45 & 1.43 & 2.88 & \textbf{3.0}\tabularnewline
2010-2017 & 1.10 & 1.68 & 2.78 & \textbf{3.58}\tabularnewline
1980-2017 & 1.38 & 1.46 & 2.84 & \textbf{3.15}\tabularnewline
\hline 
\end{tabular}
\end{table}

As discussed, the production efficiency can be related to the energy
efficiency through $\varepsilon \simeq \epsilon/\left(\lambda\tau_{d}\right)$.
This suggests that, whether it is CO$_{2}$ emissions, population,
or standard of living, it is the fractional imbalance between energy
consumption and dissipation that drives growth. A surplus enables irreversible
work to be done to make more of everything, including people, and
speeding it all up. Moreover, current efficiency levels arose from
an accumulation of prior innovations: 
\begin{equation}
\varepsilon=\int_{0}^{t}\frac{d\varepsilon\left(t'\right)}{dt'}dt'\label{eq:efficiency memory}
\end{equation}
so a conclusion might be reached that it is current and past improvements
in production efficiency that have driven current growth in emissions,
population, and standard of living. Population and emissions growth
rates have inertia because the world has memory of its past innovations.

\subsection{Emissions stabilization and climate change mitigation}

So what can be done to reverse the course of growing CO$_{2}$ emissions?
Eq. \ref{eq:Kaya new} suggests that stabilizing carbon dioxide emissions
will require the economy to decarbonize at a rate $\eta_{c}=-\lambda\varepsilon$
that is as fast as the rate of energy consumption growth $\eta_{\mathcal{E}}=\lambda\varepsilon$.
In the period 2010-2017, we observed  $\lambda\varepsilon$ to be about 2.4\%
per year (Table \ref{tab:Rates}). For a sense of what this implies,
consider that 2.4\% of the current global rate of energy consumption $\mathcal{E}\approx 20$ TW  corresponds to 480 GW. That is to say, for energy  consumption to grow at this rate without increasing carbon dioxide emissions would require over 
1 GW of new power capacity in nuclear or renewable energy to be added online each day, the approximate size of a large central power plant\footnote{The corresponding figure using $\eta_{\mathcal{E}} = 1.6\%$ directly (Table \ref{tab:Rates}), instead of the value $\lambda\varepsilon = 2.4\%$, is  320 GW, or just under 1 GW of new power capacity in renewable energy per day.}.

Furthermore, stabilized emissions do not lead to stabilized atmospheric
CO$_{2}$ concentrations, not until there is a balance with natural
uptake by the land and oceans. As a crude approximation, land and
ocean sinks are linearly proportional to the perturbation $\Delta\left[{\rm CO_{2}}\right]$
from pre-industrial concentrations of approximately 275 ppmv \cite{Wigley1983}.
More sophisticated approaches are possible that permit accurate projections
over multi-century timescales \cite{Glotter2014}. 

Nonetheless, a simple model readily lends insight to near term evolution while remaining
consistent with multiple decades of observations. Data
show that the linear sink rate $\sigma$ to the land and oceans with
respect to the decadally averaged perturbation $\Delta\left[{\rm CO_{2}}\right]$
was $2.3\pm0.5$ \% yr$^{-1}$ in the 1980s, $2.4\pm0.4$ \% yr$^{-1}$
in the 1990s and $2.2\pm0.4$ \% yr$^{-1}$ in the 2000s \cite{LeQuere2018}. There is
no inter-decadal trend and the uncertainty is greater than the variability.
Thus, an average value of $2.3\pm0.4$\% yr$^{-1}$ is assumed here.

The effect of carbon emissions on atmospheric CO$_{2}$ concentrations
can be obtained by normalizing by the atmospheric mass. Every gigaton
of emitted carbon corresponds to $0.47$ parts per million by volume
(ppmv) of increased CO$_{2}$ concentration \cite{Trenberth1981}.
The approximate balance equation is then \begin{equation}
    \frac{d\Delta\left[{\rm CO_{2}}\right]}{dt}=\kappa C-\sigma\Delta\left[{\rm CO_{2}}\right].
\end{equation}
Substituting Eq. \ref{eq:C int Y}, carbon dioxide emissions can then
be related to past economic production and the carbonization of the
energy source $c$ through the integro-differential equation
\begin{equation}
\frac{d\Delta\left[{\rm CO_{2}}\right]}{dt}=\kappa\lambda c\int_{0}^{t}Y\left(t'\right)dt'-\sigma\Delta\left[{\rm CO_{2}}\right]
\label{eq:CO2 balance GDP}
\end{equation}
Taking $\kappa$ and $\lambda$ as constants then stabilization or
reduction of concentrations at any given perturbation
value $\Delta\left[{\rm CO_{2}}\right]$ requires the following limits
on the carbonization of emissions
\begin{equation}
c\leq \frac{\sigma}{\kappa\lambda} \frac{\Delta\left[{\rm CO_{2}}\right]}{\int_{0}^{t}Y\left(t'\right)dt'} \simeq 0.26\frac{\Delta\left[{\rm CO_{2}}\right]}{W}\label{eq:c ratio}
\end{equation}
where $c$ has units of Gt C EJ$^{-1}$, $W$ has units of trillion 2010 USD and the numerical coefficient $\sigma/\left(\kappa\lambda\right)\simeq 0.26$ has units of 
$({\rm Gt \, C})\times({\rm trillion \, 2010 \, USD})\times ({\rm ppmv}^{-1})\times({\rm EJ}^{-1})$.

Recent values for $c$ are close to 0.017 Gt C EJ$^{-1}$, and despite a recent surge in renewables, $c$ is not rapidly declining \cite{Jackson2018}.
As shown in Table \ref{tab:Rates-C}, the annual decarbonization rate $\eta_c$
in recent years is just 0.36\% yr$^{-1}$ and the correspondence between
CO$_{2}$ emissions $C$ and cumulative global production $W$ expressed as $C/W=\lambda c$ has effectively been unchanged over the past four decades. Based on these observations, to develop an intuition for constraints on the future, we take
the baseline assumption that $\lambda c$ will remain fixed. In this case, Eq. \ref{eq:CO2 balance GDP} implies a straightforward proportionality relating the world cumulative production $W$ and the equilibrium value for the concentration perturbation $\Delta\left[{\rm CO_{2}}\right]_{\rm eq}$. 
Setting $d\Delta\left[{\rm CO_{2}}\right]/dt=0$ it follows that:
\begin{equation}
W=\frac{\sigma}{\kappa\lambda c} \Delta\left[{\rm CO_{2}}\right]_{\rm eq}\simeq  15.4 \Delta\left[{\rm CO_{2}}\right]_{\rm eq}\label{eq:Sum GDP CO2}
\end{equation}
Here, the numerical coefficient $\sigma/\left(\kappa c\lambda\right)\simeq 15.4$ is obtained from the years 1980 to 2010 and has units of trillion 2010 USD ppmv$^{-1}$. Values of this coefficient for different periods are shown in Table \ref{tab:WDCO2}. 

\begin{table}[ht]
\centering
\caption{\label{tab:WDCO2}Average values of the scaling $W/\Delta\left[{\rm CO_{2}}\right]_{\rm eq} = \sigma/\left(\kappa c\lambda\right)$ defined by Eq. \ref{eq:Sum GDP CO2}
for various time periods, in trillion 2010 USD ppmv$^{-1}$}

\begin{tabular}{l|c|c|c|c}
\hline 
Period & 1980-1990 & 1990-2000 & 2000-2010 & 1980-2010 \tabularnewline
\hline 
$\sigma/\left(\kappa c\lambda\right)$ & 15.0 & 16.3 & 14.9  & 15.4 \tabularnewline
\hline 
\end{tabular}
\end{table}

Comparing Eq. \ref{eq:Sum GDP CO2} with Eq. \ref{eq:E lambda W} we obtain  
\[
\Delta\left[{\rm CO}_{2}\right]_{\rm eq}\simeq \frac{\text{\ensuremath{\kappa c}}}{\sigma\tau_{d}}G
\]
Effectively, at equilibrium with land and ocean sinks, civilization's
combustion garbage heap in the form of an atmospheric CO$_{2}$ perturbation
is linearly proportional to how far thermodynamically it has stretched
itself away from the environmental base state of $G=0$. Both rise
as our collective historical achievement.

\begin{figure}[ht]
\includegraphics[width=12cm]{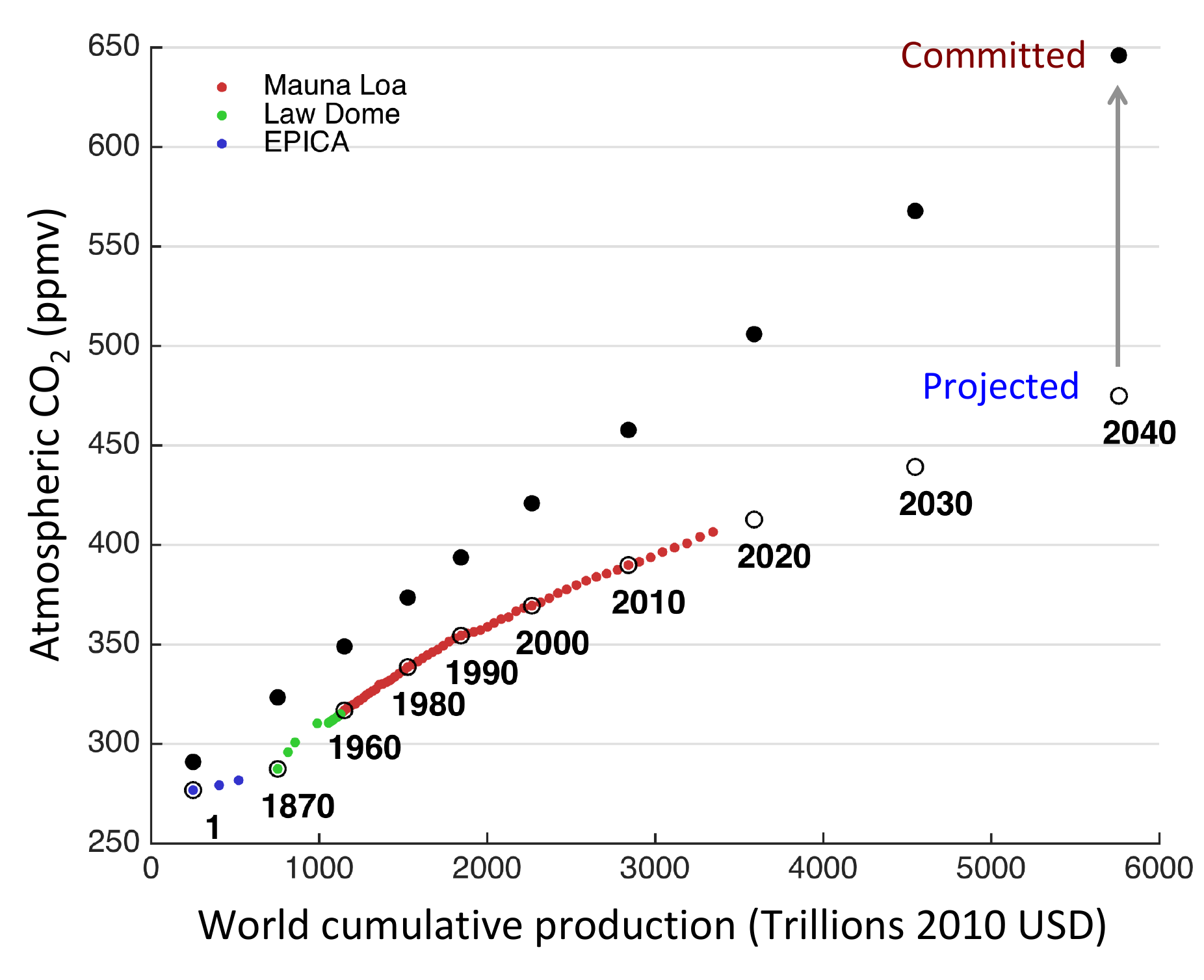}

\caption{Historical reconstructions of world cumulative production $W$ and
atmospheric CO$_{2}$ concentrations with projections assuming $\eta_{c}=0$
and $\eta_{\mathcal{E}}=\eta_{\mathcal{E}}\left(2017\right)=2.4\%$yr$^{-1}$
(circles) and corresponding stabilization concentrations from Eq.
\ref{eq:Sum GDP CO2}. The halving time between predicted and committed
is about 30 years. Concentration data includes
flask samples from Mauna Loa \cite{KeelingWhorf2005} and Antarctic
ice core data \cite{Etheridge1996,Fluckiger2002}. } \label{fig:CO2andW}
\end{figure}

Figure \ref{fig:CO2andW} shows the relationship between the world
cumulative production $W$ and CO$_{2}$ concentrations, both for
the past 2000 years and for the future assuming no future decarbonization
and that energy consumption persists at its current growth trend of
2.4\% yr$^{-1}$, a conservative estimate perhaps given that it implies
an equivalent rate of world GDP growth (Eq. \ref{eq:GDP growth efficiency}). That is to say, the open circles for each future date correspond to the non-equilibrium values for CO$_{2}$ concentrations derived from Eq. \ref{eq:CO2 balance GDP} assuming that $W$ grows at the constant rate $\eta_W=\lambda \varepsilon$. In addition, 
for each value of $W$ the set of solutions is provided for the CO$_{2}$ concentration at which emissions stabilize with land and ocean sinks by applying Eq. \ref{eq:Sum GDP CO2}, namely the equilibrium concentration to which civilization is committed even for the mathematically extreme case that $W$ were to remain constant (in other words, not only zero GDP growth, but effectively zero inflation-adjusted production).

The equilibrium CO$_{2}$ concentration is approached
asymptotically with timescale $\tau_{{\rm CO_{2}}}=1/\sigma$, so
that the difference is halved in about 30 years. For example, Figure
\ref{fig:CO2andW} shows that stabilizing concentrations at a nominal
value of 350 ppm would require that the current world cumulative production
shrink by two thirds to a value not seen since 1960. 

It is probably safe to assume that civilization will not willingly
engage in such drastic pruning. 
Looking to the future, Figure \ref{fig:CO2andW} shows
that without rapid decarbonization, we have already committed ourselves
to CO$_{2}$ concentrations above $500$ ppmv, well in excess of
the 450 ppmv threshold that has been deemed ``dangerous'' \cite{HansenDangerous2007}.
At current growth rates, the commitment is to a doubling of pre-industrial
levels by 2030, and to levels close to $650$ ppmv by 2040.

\section{Conclusions}

This article identifies a persistent relationship between global energy
consumption and cumulative economic production. It implies that a surprisingly 
simple description of the human system is sufficient to explain  past
global trends and make robust projections of the aggregated world economy and its waste products.
Humanity grows when more energy is available than required for 
daily needs. Then work can be done not just for sustenance but for
expansion. Because current sustenance demands emerge from past growth, inertia
plays a much more important role in determining future societal and
climate trajectories than has been generally acknowledged, particularly
in the physically unconstrained models that are widely used to link
the economy to climate \cite{tol2009economic,Nordhaus2013DICE}. We have
accumulated over history a long series of innovations in efficiency that continue to propel us forward. Without forgetting these advances, we will maintain a continued ability to expand our interface with the primary resources we consume. 

Eventually, of course, the interwoven networks of civilization will
unravel and emissions will decline, whether it is through depletion
of resources, environmentally forced decay or -- as demonstrated
recently -- pandemics \cite{LeQuere2020}. But the cuts will have
to be deep, continuous, and cumulative to overcome the tremendous
accumulated growth we have sustained up to this point. 

The formulations presented here are intended to help constrain the problem by reducing the number of available targets that can reasonably be expected to
lead to avoidance of extreme climate change. Notably, gains in energy efficiency play a critical role in enabling increases in population and prosperity, and in turn growth of energy demands and
carbon dioxide emissions, contrary to what would reasonably be assumed if civilization did not grow \cite{Koomey1998, Raupach2007, Andreoni2016}. What
seems to be required is a peculiar dance between reducing the production
efficiency of civilization while simultaneously innovating new technologies
that move us away from combustion. 

The relationships identified all stem mathematically from the falsifiable
identify $\mathcal{E}=\lambda W$ where $W=\int_{0}^{t}Y\left(t'\right)dt'$.
While the specific value of $\lambda$ that was identified is 5.9$\pm$0.2
gigawatts per trillion 2010 US dollars, what matters from the standpoint
of calculating trends is that the ratio $\mathcal{E}/\int_{0}^{t}Y\left(t'\right)dt'$
is a constant to within observational uncertainty. Further theoretical work is required to link the relationship to more traditional macroeconomic modeling frameworks.  Continued observations will provide a useful check on its validity. Any evidence of a sustained downward trend in $\lambda$ help to pinpoint where there is a decoupling of economic production from civilization's metabolic needs.

\section*{Acknowledgments}
This work was supported by the United Kingdom Economic and Social Research Council whose views it does not represent. 

\nolinenumbers

%
%
%

\bibstyle{plos2015}
\bibliography{References}

\appendix

\section{Calculation of cumulative production}
\label{S1_Appendix}

Market exchange rate estimates of $Y_{i}$, inflation-adjusted to
``real'' constant year 2010 dollars, are available from the World
Bank and the United Nations for the years between 1970 and 2017 \cite{WorldBank2019,UNstats}.
Estimates of real GDP adjusted for purchasing power parity (PPP) 1990
USD are available for each year between 1950 and 1992, and in larger
intervals extending back to 1 CE \cite{Maddison2003}. To calculate
$W$ these estimates are converted to market exchange rate MER inflation-adjusted
2010 values. For the time period between 1970 and 1992 for which concurrent
MER and PPP statistics are available, the mean inflation-adjusted
ratio PPP/MER is $\kappa = 1.205$ with no clear trend.

A historical reconstruction of the annual global GDP is obtained by
dividing the Maddison PPP values by $\kappa$ between
1 C.E. and 1970 C.E, applying a cubic spline between sparse data points
to obtain annual values, and using World Bank statistics for more
recent years \cite{WorldBank2019}. The value of world cumulative
production $W$ is then 
\begin{equation}
W\left(t\right)=W\left(1\right)+\sum_{1}^{t}Y\left(t\right)\label{eq:CfromGWP}
\end{equation}
where $W\left(1\right)$ refers to total accumulated world cumulative
production to date in 1 C.E. To obtain a value for $W\left(1\right)$,
it is assumed that $W$ and world population grew equally fast at
that time. Available statistics suggest a population in ca. 1 C.E.
\cite{USCensus} that was 170 million and
growing by 10 million every hundred years, at a rate of $\eta_{pop}=$0.059
\% per year. The estimated value for the real MER GDP in 1 C.E. is
0.147 trillion 2010 USD. Assuming that civilization population and
wealth grew at the same rate, i.e., $\eta_{pop}=\eta_{W}$, then from
Eq. \ref{eq:WintY} it follows that $W\left(1\right)=250$ trillion
2010 USD.

One criticism might be that MER dollars should be adjusted to PPP
dollars \cite{Cullenward2010} since market exchange rates fail to
account for differences in how people in different countries value
equivalent baskets of goods. One rebuttal has been that such equivalents
do not exist because different cultures value goods differently and
that any discrepancies tend to diminish over time with a half life
of three to five years due to the pressures of international and domestic
trade \cite{Rogoff1996}. In the case of the work here, there is
another counter-argument which is that there is no intent to address
short-term inequalities between nations, only the global sum of all
of civilization and its evolution over they long-run. Effectively,
there is only one ``basket of goods'', and that is humanity taken
as a whole, including all its social and physical networks.

Rates of global primary energy consumption from all sources $\mathcal{E}$
are available from the U.S. Department of Energy (DOE) Energy Information
Administration (EIA) for the time period 1980 to 2016 and from British
Petroleum between 1965 and 2017 \cite{AER2011,BritishPetroleum2018}.
Rates of global primary energy consumption and production provided
by the EIA have a mean ratio of 99.83\% so here it is assumed that
the two are equivalent.


\end{document}